\documentclass[twocolumn,prl]{revtex4}

\usepackage{amssymb}
\usepackage{amsmath}
\usepackage{graphicx}
\usepackage{psfrag}
\usepackage{verbatim}

\begin{document}

\title{Laser cooling of a nanomechanical resonator mode to its quantum
  ground state}

\author{I. Wilson-Rae$^{1,2}$, P. Zoller$^{3}$, and A. Imamo\=glu$^{2}$}

\affiliation{$^1$ Department of Physics, University of California,
Santa Barbara, CA 93106}

\affiliation{$^2$ Institute of Quantum Electronics, ETH
H\"onggerberg HPT G12, CH-8093 Z\"urich, Switzerland}

\affiliation{$^3$ Institut f\"ur Theoretische Physik,
Universit\"at Innsbruck, A-6020 Innsbruck, Austria}

\date{\today}

\begin{abstract}
We show that it is possible to cool a nanomechanical resonator mode to
its ground state. The proposed technique is based on resonant laser
excitation of a phonon sideband of an embedded quantum dot. The
strength of the sideband coupling is determined directly by the
difference between the electron-phonon couplings of the initial and
final states of the quantum dot optical transition. Possible
applications of the technique we describe include generation of
non-classical states of mechanical motion.
\end{abstract}

\maketitle

In contrast to bulk semiconductors where lattice vibrations
form a reservoir with a continuous spectrum, nano-scale
semiconductor structures support sharp mechanical resonances with
very high quality factors and frequencies approaching 1
gigahertz\cite{Roukes01}. Unlike optical resonators, the thermal
occupancy of these mechanical resonators is well above unity, even
when they are cooled to sub-Kelvin temperatures. Presence of
thermal noise hinders the study of coherent quantum dynamics in
these mechanical systems \cite{Armour02} and limits their
applications to precision measurements \cite{Cleland02}.  Here, we
propose a technique that would realize laser cooling of a
nano-mechanical resonator mode to its motional ground-state where
quantum effects will be manifest. The basic idea is to use light
scattering and electron-phonon interactions\cite{Mahan}, such as
deformation potential coupling\cite{Ridley}, in an embedded
quantum dot to manipulate a discrete mode of lattice vibrations.

The semiconductor beam structure with an embedded quantum dot (QD)
that we analyze is shown in Fig.~1. An infinite length beam
supports four phonon branches without an infrared cutoff: two
bending branches (in-plane bending and flexural) with quadratic
dispersion relations, and a torsional and a compression branch
with linear dispersion relations\cite{Landau,Graff,Cross01}. For a
finite beam of length $L$ attached to supports via abrupt
junctions, these branches at long wavelengths will develop a series of
sharp resonances corresponding to the different harmonics
\cite{Cross01}. In a structure where the thickness of the
beam ($d$) is smaller than its width ($b$) (Fig.~1), the
lowest-energy resonance corresponds to the fundamental flexural
(vertical bending) mode ($\hbar \omega_0$). This will constitute
the resonator mode we intend to address and cool to the ground
state. For flexural modes with $\omega_0 \sim
10^9\textrm{s}^{-1}$, quality factors ($Q$) exceeding $20,000$
have been measured \cite{Cleland01}. Though we focus on a bridge
geometry (Fig.~1), the analysis of a cantilever structure is
completely analogous.

The presence of a zero-dimensional (anharmonic) emitter that can be
resonantly excited by a laser, has a spontaneous emission broadened
optical transition, and interacts with the acoustic modes is necessary
for achieving ground-state cooling. Our discussion will primarily
focus on self-assembled InAs QDs \cite{Petroff01} which can satisfy these
requirements and can be modeled as two level systems consisting of the
empty QD state ($|g\rangle$) and the fundamental exciton state
($|e\rangle$)\cite{Becher01}. We envision an ideal scenario where
processing leaves only one QD inside the beam at a specific
location. In assuming a natural line-width limited QD we are ignoring
the effect of the surface on the electronic properties of the
dot. This is warranted provided the dephasing and/or non-radiative
recombination rates arising from the proximity to semiconductor/air
interfaces are negligible compared with the spontaneous emission rate.

\begin{figure}[b]
\vspace{-1.3cm}
\psfrag{a}{}
\psfrag{J}{$\,${\scriptsize II}}
\psfrag{l}{$L$}
\psfrag{b}{$\!\!b$}
\psfrag{x}{${}^{x}$}
\psfrag{y}{${}_{y}$}
\psfrag{z}{${}^{z}$}
\centerline{\includegraphics[width=9cm]{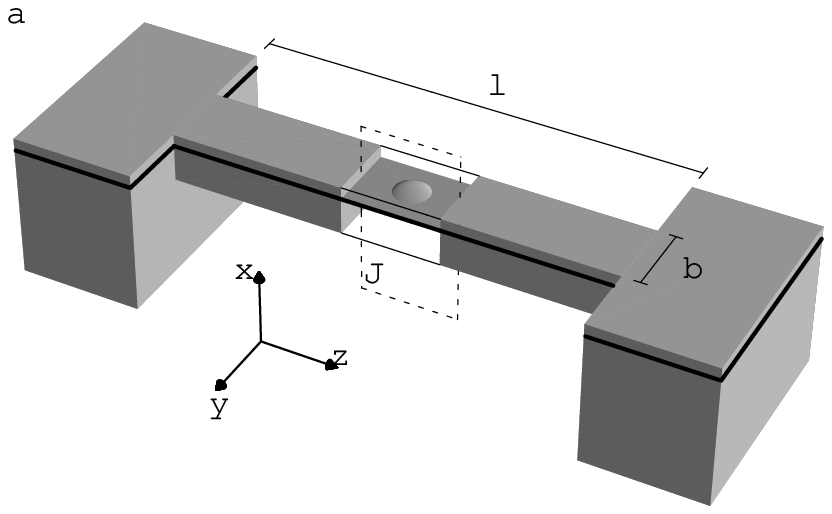}}

\vspace{-1.3cm}
\psfrag{i}{\scriptsize I}
\psfrag{j}{\scriptsize II}
\psfrag{k}{\scriptsize III}
\psfrag{q}{\scriptsize $\!\!\!$QD}
\psfrag{d}{$\!\! d$}
\psfrag{B}{}
\psfrag{w}{\scriptsize WL}
\centerline{\includegraphics[width=7.5cm]{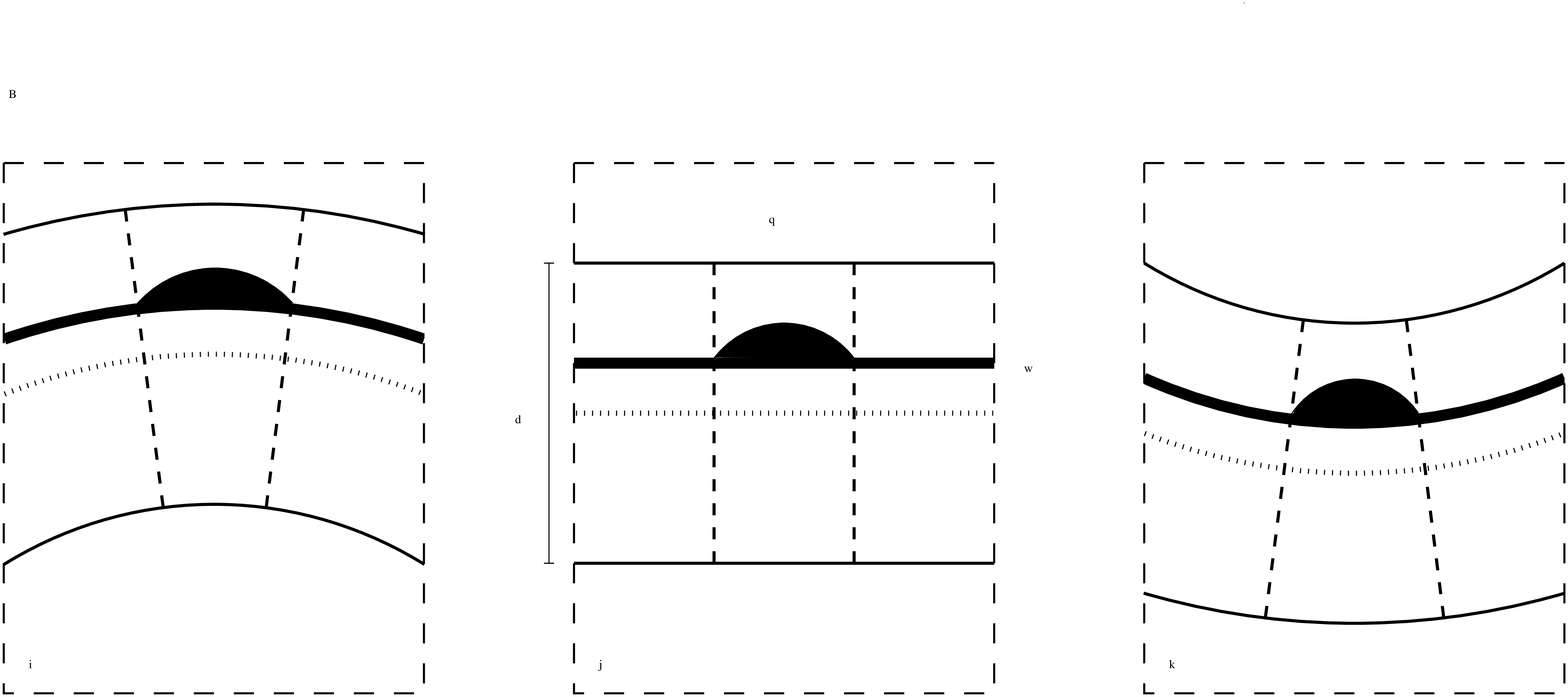}}
\caption{Top: Schematic diagram of a GaAs bridge of length $L$, width
$b$ and thickness $d$ (see II below), with an embedded lens shaped
InAs QD (WL denotes the wetting layer). Bottom: Vertical
cross-sections through the axis of the bridge illustrating the
deformations suffered by a small neighborhood of the QD, when the
beam is bended slightly in the vertical direction ($x$). I and III
correspond respectively to positive and negative deflection and II
shows the equilibrium configuration. Each volume
element parallel to the axis undergoes a simple extension or
compression\cite{Landau}.}
\end{figure}

We observe in Fig.~1 that flexion induces extensions
and compressions in the structure\cite{Landau}. This longitudinal
strain will modify the energy of the electronic states confined in
the QD through deformation potential coupling\cite{Ridley}. The
resulting interaction between a localized carrier and a discrete
flexural phonon mode will lead to the appearance of sidebands in
optical spectra, originating from phonon assisted photon
absorption or emission processes \cite{Mahan}. By tuning a laser
field into resonance with the first lower energy (red) sideband of
a QD optical transition, we can ensure that absorption of a laser
photon will be accompanied by the removal of a phonon from the
system\cite{ionreview,ion}.

Before proceeding we note that {\sl sideband cooling} has been used to
cool trapped ions to their motional ground state
\cite{ionreview}. Despite the analogy, the physics of the two systems
are different. Unlike the case of trapped ions \cite{ion} and in
marked contrast to optomechanical cooling schemes \cite{Vitali03}, the
photon recoil plays no role in the laser cooling of a nanomechanical
mode using an embedded QD\footnote{
We have estimated Doppler effects and radiation pressure in our system
and find them to be negligible.
}. In the latter system, the (oscillator)
strength of the sidebands is determined directly by the difference
between the electron-phonon couplings of the initial and final states
of the optical transition\cite{Mahan}.

The fact that the modes of interest correspond to acoustic phonons
with typical wavelengths $\lambda_p \sim L \gg b, d$ (Fig.~1) orders
of magnitude larger than the lattice constant makes thin rod
elasticity theory an excellent approximation to describe the phonons
inside the beam\cite{Landau,Graff,Cross01}. The dominant
electron-phonon interaction mechanism in our structure will be
deformation potential coupling\footnote{
The structure can be processed so that $\{x,y,z\}$ (Fig.~1) are
oriented along the principal crystal axes. This choice
implies that piezoelectricity will only couple the QD to
torsion\cite{Ridley,Landau}. We have bounded the corresponding
Huang-Rhys parameters and find them to be negligible.
}
\begin{equation}\label{DP}
H_{\mathrm{DP}} = \int \mathrm{d} \bar{r}^3 \left[ D_c
\hat{\rho}_{\rm{el}} \left( \bar{r} \right) - D_v \hat{\rho}_{\rm{h}}
\left( \bar{r} \right) \right] \nabla \cdot \hat{\bar{u}} (\bar{r})\,.
\end{equation}
Here, we have introduced the electron (hole) density operator
$\hat{\rho}_{\rm{el}}$ ($\hat{\rho}_{\rm{h}}$) and the corresponding
deformation potential constants $D_c$ ($D_{v}$). The vector field
operator $\hat{\bar{u}}(\bar{r})$ corresponds to the displacement in
Lagrangian elasticity\cite{Landau}. We neglect the difference between
the elastic properties and deformation potentials of the strained InAs
QDs and the surrounding GaAs matrix\footnote{
The smallness of the QD dimensions implies that for phonons
characterized by $\lambda_p$ a QD-phonon interaction $\propto \nabla
\cdot \hat{\bar{u}} (\bar{r})|_{\bar{r}=\bar{r}_{\mathrm{D}}}$ can be
justified without resorting to this approximation\cite{IWPZAI} --
$\bar{r}_{\mathrm{D}}$ is the QD position.
}. 

To proceed we decompose $\hat{\bar{u}}(\bar{r})$
in terms of a set of orthonormal modes.  A possible set
consists of discrete modes localized in the beam
($\{\bar{u}_{m}(\bar{r}) \}$) satisfying clamped boundary conditions
at the junctions, and a continuum of modes localized in the supports
($\{\bar{u}_{\tilde q}(\bar{r}) \}$) satisfying free boundary
conditions. In this representation the beam modes only interact with
support modes\cite{Viviescas03} and for long wavelengths ($\lambda_p$)
these interactions are weak, making the $\{ \bar{u}_{m}(\bar{r}) \}$
correspond to the resonances\cite{Cross01}. As we are just interested
in the dynamics of the fundamental mode (described by
$\bar{u}_{0}(\bar{r})$, $b_0$), we single it out and diagonalize all
the terms in the phonon Hamiltonian which do not involve $b_0$, by
transforming the rest of the modes to a new continuum
$\{\bar{u}_{q}(\bar{r}) \}$. Thus we obtain a representation
consisting of the discrete resonator mode weakly coupled to a
reservoir of ``background'' modes $\{b_q\}$.

We subsequently restrict (\ref{DP}) to the two level model for the QD
and integrate over $\bar{r}$ to obtain a coupling term between the QD
and the resonator mode of the form: $\hbar \tilde \omega_0 \eta
|e\rangle \langle e | ( b_0^{\vphantom\dagger} + b_0^\dagger )$, where
$\tilde\omega_0$ is the bare frequency associated to the discrete
resonator mode. Within thin rod elasticity, $\nabla \cdot
\bar{u}(\bar{r})$ for a bending mode is well approximated by $(2
\sigma -1) x \frac{\partial ^2 X}{\partial z^2}$, where $X (z)$
(Fig.~1) is the deflection of the center-of-mass of the cross-section
of the beam, $\sigma$ is the Poisson ratio \cite{Landau} and the
coordinate origin is at the midpoint. We use this relation, that
$\lambda_p$ is much larger than the typical QD size\cite{Becher01},
and that the differences between the average positions of the hole,
the electron and the center of mass of the QD can be neglected, to
obtain:
\begin{equation}\label{eta}
\eta^{2} = \frac{3 \sqrt{6} \left( 1 - 2 \sigma \right)^2
\!(D_c-D_v)^2}{\pi^2 \hbar \rho c_T^3 \left( 1 + \sigma \right)^{3/2}}
\,\, \frac{x_D^2 L}{d^4 b} \,\, \frac{{X''_0}^2 \left( k_0 z_D
\right)}{\left( k_0 L/\pi \right)^2}\,.
\end{equation}
To the extent that one neglects the difference between the bare
frequency $\tilde\omega_0$ and the actual resonance $\omega_0$,
$\eta^2$ corresponds to the Huang-Rhys parameter of the resonator
mode. Here $\rho$ and $c_T$ are the density and transverse speed of
sound for the material of the beam, and $x_D$ and $z_D$ are the QD
coordinates (Fig.~1). The mode function normalized to the length and
the wave-vector of the resonator mode are denoted by $X_0 \left( k_0 z
\right)$ and $k_0$, respectively.  The first factor in Eq.~(2) is the
square of a characteristic length of the material ($2$~nm for
GaAs). The second factor gives the dependence with the dimensions of
the structure and the placing of the QD in the vertical direction. The
last factor will depend on whether we have a cantilever or a bridge,
and for the latter is of order unity in a neighborhood of the
midpoint. Typical parameters (see below) yield $\eta \sim
0.06$.

The Hamiltonian describing our system includes terms corresponding
to the interactions between the QD, the resonator mode, the
background phonon modes, and the electromagnetic field. To study
laser cooling we perform a canonical transformation that
eliminates the QD-resonator mode coupling term\cite{Mahan}. After
additional coordinate and momentum  shifts applied to the phonon
modes, we obtain the transformed Hamiltonian:
{\setlength\arraycolsep{2pt}
\begin{eqnarray} \label{ham} H & = & H_0
-\hbar \delta \frac{\sigma_z}{2} + \hbar \!\left[\frac{\Omega}{2}
\sigma_{+} B^\dagger + \sum_k  g_k \sigma_{+} B^\dagger a_k \right.
\nonumber\\ & & \left.\!\! +\, \frac{\sigma_{z}}{2} \sum_q
\lambda_q b_q +  \left(b_0^{\vphantom\dagger} + b_0^\dagger
\right) \sum_q \zeta_q b_q +  \textrm{h.c.} \right]\!,\,\,
\end{eqnarray}}\noindent
where
$H_0= \hbar [ \tilde\omega_0 b_0^\dagger b_0^{\vphantom\dagger} +
\sum_q \omega_q b_q^\dagger b_q^{\vphantom\dagger}+ \sum_k \left(
\omega_k - \omega_L \right) a_k^\dagger a_k^{\vphantom\dagger} ]$
and the QD operators are expressed in terms of Pauli matrices
(i.e. $|e \rangle \langle e| = 1/2 ( 1 + \sigma_z)$). We have also
defined the polaron operator $B = e^{\eta \left(
b_0^{\vphantom\dagger} - b_0^\dagger \right)}$, the detuning of
the laser from the fundamental exciton line $\delta$ and its
frequency $\omega_L$ and Rabi frequency $\Omega$.  The background
phonon modes are characterized by annihilation operators $b_q$ and
their couplings to the QD (resonator mode) $\lambda_q$
($\zeta_q$). The radiation field is characterized by
annihilation operators $a_k$ and their couplings to the QD $g_k$.

The role of $\eta$ in our system is equivalent to the one played by
the Lamb-Dicke (LD) parameter in the trapped-ion
system\cite{ionreview,Cirac92}. Even though the underlying physics is
different, we will refer to the LD regime as the one in which
$\eta^2_{\rm eff}\equiv \eta^2 (\langle b_0^\dagger b_0^{\vphantom\dagger}
\rangle +1 ) \ll 1$. We focus on this limit since it is the one of
interest for ground state cooling and is well satisfied for realistic
structures provided the ambient temperature is low enough.  In this
regime, both cooling and heating can take place via two paths, where
the modification of the motional state of the phonon mode takes place
in the laser absorption and spontaneous emission process, respectively
(Fig.~2). In contrast to the trapped ion system, these two paths are
indistinguishable for the QD-beam structure and exhibit quantum
interference.

In this LD regime it is possible to derive a rate equation describing the
cooling dynamics of the resonator mode. To this effect we 
first eliminate the radiation field and subsequently the
``background'' phonons to obtain a master equation for the
QD-resonator system. The first step, that involves the standard
Born-Markov approximation, leads to a Liouvillian for the
``QD-resonator-background'' system that -- aside from terms involving
the ``background'' -- closely resembles the analogous result for a
trapped-ion\cite{Cirac92}. The only difference -aside from a trivial
interchange of the normal coordinate and momentum of the resonator
mode- lies in the dissipative part: $\Gamma/2(2\sigma_- B \rho
B^{\dagger} \sigma_+ - \rho \sigma_+\sigma_- - \sigma_+\sigma_-\rho)$,
where $\rho$ is the density matrix. The crucial point is that $B$ does
not depend on the electromagnetic modes, in marked contrast to the
momentum shift operator that plays the analogous role for a
trapped-ion.  This gives rise to the interference effects already
discussed. 

\begin{figure}
\psfrag{g}{$g_k$}
\psfrag{d}{$\!-\hbar \delta$}
\psfrag{w}{$\hbar \omega_0$}
\psfrag{p}{$\!\frac{\Omega}{2}$}
\psfrag{h}{$v_{\rm S}$}
\psfrag{o}{$v_{\rm L}$}
\psfrag{r}{$\!\!\!|g,n-1\rangle$}
\psfrag{s}{$\!\!\!|g,n\rangle$}
\psfrag{t}{$\!\!\!|g,n+1\rangle$}
\psfrag{u}{$\!\!\!|e,n-1\rangle$}
\psfrag{v}{$\!\!\!|e,n\rangle $}
\psfrag{x}{$\!\!\!|e,n+1\rangle$}
\psfrag{k}{$\!v_{\rm S} \equiv \!\eta g_k \sqrt{n}$}
\psfrag{l}{$\!v_{\rm L} \equiv \!-\eta \frac{\Omega}{2} \sqrt{n}$}
\centerline{\includegraphics[width=8.5cm]{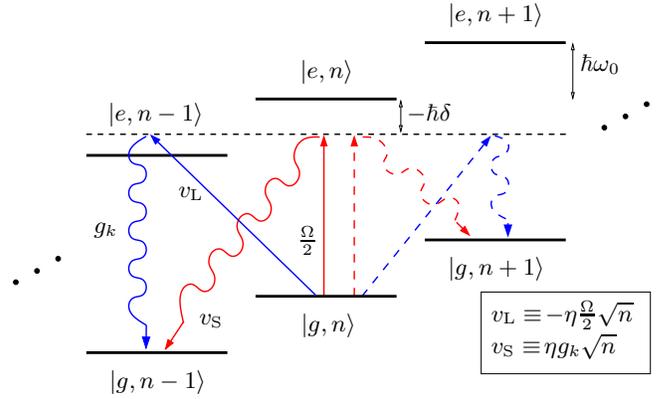}}
\caption{ Energy level diagram of the QD coupled to
the resonator mode for perturbative Rabi frequency. $|g,n \rangle$
($|e,n \rangle$) denote the state where the QD is in the
ground (exciton) state and the resonator mode has $n$ phonons.
The cooling (heating) cycle consists of the excitation of the
QD and subsequent spontaneous photon emission, and is
denoted by solid (dashed) lines. The blue (red) path corresponds
to the case where phonon number is changed during the laser
absorption (spontaneous emission). If the laser is tuned between
the red sideband and the main line these two
paths interfere constructively for cooling and destructively for
heating (note the relative phase of $\pi$ between $v_{\rm S}$ and
$v_{\rm L}$).}
\end{figure}

The background will exhibit sharp resonances corresponding to the
other beam modes ($\bar{u}_{m}(\bar{r})|_{m \neq 0}$). It can
be shown that provided these resonances are well detuned from
$\omega_0$ and $2\omega_0$ (see \footnote{
This can be satisfied by processing the structure so that $b/d > 25/9$
(Ref.~\cite{Landau,Graff,Cross01}).
}), the effect of the couplings $\lambda_q$ in Hamiltonian (\ref{ham})
on the cooling dynamics of the resonator mode is higher order in the
small parameters $\eta^2$ and $d/L$
(Ref.~\cite{IWPZAI}). Hence we neglect these couplings in our
derivation of a master equation for the QD-resonator system. The
weakness of the interaction between the resonator and the
supports\cite{Cross01} allows us to apply the Born-Markov
approximation to the remaining couplings $\{\zeta_q\}$. This yields
three distinct contributions: a Hamiltonian part that shifts
$\tilde\omega_0$ to $\omega_0$, a ``counter-rotating'' part that to
lowest order in $1/Q$ will not contribute to the rate equation, and a
``rotating wave'' part describing the damping of the resonator\cite{Walls}.

We proceed to apply the LD approximation following
Ref.~\cite{Cirac92}. This involves expanding all the polaron
exponentials $B$ up to second order in $\eta$ and adiabatically
eliminating the QD. The damping of the resonator can be included
using an ``LD-large $Q$'' approach in which $1/Q$ is treated on the
same footing as $\eta^2$. The resulting rate equation describing the
populations $P_n$ of the energy levels of the resonator mode is
\begin{align}\label{rate}
&\dot{P}_n = \left[ \eta^2 A_+ + \omega_0 \textstyle{\frac{n \left(
   \omega_0 \right)}{Q}} \right] \left[n P_{n-1} - \left( n+1 \right)
   P_ n \right] \nonumber\\ & + \, \left[ \eta^2 A_- + \omega_0
   \textstyle{\frac{n \left( \omega_0 \right) + 1}{Q}} \right]   \left[\left( n+1 \right) P_{n+1} - n P_n \right],
\end{align}
with $n(\omega_0)=1/[e^{\omega_0/k_B T}-1]$. The first term
corresponds to heating and the second one to cooling. There are
contributions proportional to $1/Q$ describing the damping arising
from coupling to the supports and contributions which involve the
scattering of laser light given by $\eta^2 A_\pm =
2\eta^2\mathrm{Re}\{\sum_{\nu}\mathbf{A^{\pm}}_{0\nu}\langle
\sigma_\nu\rangle_{\mathrm{SS}}\}$, with
\begin{equation}
\mathbf{A^{\pm}} =\textstyle{\frac{\Gamma}{2}} \mathbf{C}\, +\,
\left(\textstyle{\frac{\Omega}{2}} \, \mathbf{R}^y -
\textstyle{\Gamma} \mathbf{C} \right) \cdot \left( \mp i\omega_0
\mathbb{I}\! - \mathbf{L} \right)^{-1} \cdot \left(
\textstyle{\frac{\Omega}{2}} \, \mathbf{R}^y - \textstyle{\Gamma}
\mathbf{C}\right).
\end{equation}
%
We have used for the state space of the QD a basis consisting of
the Pauli matrices ($\sigma_\nu$ with $\nu > 0$) and the identity
($\sigma_0$), and defined: $R^y_{\mu \nu} = \frac{1}{2}
\mathrm{Tr}_{{}_{\mathrm{QD}}}\left(\sigma_\nu \sigma_y \sigma_\mu
\right)$, $C_{\mu \nu} = \frac{1}{2} \mathrm{Tr}_{{}_{\mathrm{QD}}}
\left( \sigma_\mu \sigma_- \sigma_\nu \sigma_+ \right)$ and $L_{\mu
\nu} = \frac{1}{2} \mathrm{Tr}_{{}_{\mathrm{QD}}} (\sigma_\mu
\mathcal{L}^{{}^{\mathrm{QD}}} \sigma_\nu)$. Here
$\mathcal{L}^{{}^{\mathrm{QD}}}$ and $\langle
\sigma_\nu\rangle_{\mathrm{SS}}$ are the Liouvillian and steady state
expectation values corresponding to the optical Bloch equations for
the QD\cite{Cohen}. We focus on $\delta <0$ for which there is
a net cooling rate $\eta^2 W \equiv \eta^2 (A_- - A_+)>0$.

As we start from thermal equilibrium, initially the number of phonons
is given by $n_i = n(\omega_0)$. When steady state is reached, the
final phonon number $n_f$ will be given by $n_f= (n_i + \eta ^2 Q
A_+/\omega_0)/(1 + \eta ^2 Q W/\omega_0)$.  Therefore there is a
threshold value of $n_i = \min \{ A_+/W \}|_{\Omega,\delta}$ below
which laser cooling does not work (Fig.~3), and $\eta^2
Q >> 1$ is required to obtain appreciable cooling.

There are two contributions to $n_f$: the first one is proportional to
$n_i$ and the second one proportional to $A_+$.  This leads to two
regimes depending on which contribution dominates the behavior of
$n_f$ (Fig.~3). In the first ``large $n_i$-small $\eta^2 Q$'' regime
the dominant heating mechanism is the coupling to the supports. In the
second ``small $n_i$-large $\eta^2 Q$'' regime heating is dominated by
the scattering of laser light and the optimal value of $n_f$
($\tilde{n}_f$) is given by $\tilde n_f \simeq (\Gamma/4\omega_0)^2$
(Fig.~3). We find that in both regimes $\Gamma/\omega_0 < 1$ is a
necessary condition for ground state cooling and $\tilde n_f \lesssim
n_i/\eta^2 Q + (\Gamma/4\omega_0)^2$.  Typical parameters we envisage
are: $L=950$nm, $d=30$nm (which correspond to $\omega_0 \approx 1.2
\times 10^{9}$s$^{-1}$), $b=85$nm, $x_D=d/4$, $z_D=0$, $Q=3 \times
10^{4}$ and $\Gamma = 3 \times10^8{\rm s}^{-1}$\cite{Becher01}. These
lead to $\tilde n_f = 0.1$ for a sample held at $T = 0.1K$ (Fig.~3).

We note that it is possible to realize an all optical measurement of
the final temperature. The basic idea is to take two lasers, a first
one for cooling with the desired parameters, and a second one for
measuring with zero detuning (frequency $\omega_M$) and perturbative
Rabi frequency (e.g. $\Omega_M = \Gamma / 10$). While this second
``probe'' laser is always on, the first ``cooling'' laser interacts
with the QD only during time windows of duration $t_c$ (cooling) that
alternate with measurement windows of duration $t_m \ll t_c$. As for
realistic parameters optimal cooling occurs at $\Omega \sim \omega_0$,
when analyzing the cooling windows the ``probe'' laser can be
neglected. The experimentally measured quantities are the photo-counts
$N_-$ (blue photons) and $N_+$ (red photons) around frequencies
$\omega_M + \omega_0$ and $\omega_M - \omega_0$ respectively, during
the time-intervals when the ``cooling'' laser is off. The cooling and
heating rates satisfy $A_\mp(\delta) = A_\pm(-\delta)$. Hence if we
assume that there are no background photons, it follows that in the
above scenario the occupancy will be given by $\langle b_0^\dagger
b_0^{\vphantom\dagger} \rangle = N_- / (N_+ - N_-)$.

We have focused on laser cooling the fundamental resonator mode to its
ground state. However we want to stress that one of the main
contributions of this letter is to establish a close analogy between
this mesoscopic condensed-matter system and a single trapped ion in
the LD regime\cite{ionreview,Cirac92}. In fact, the analogy extends to
cavity-QED; i.e. a single atom in a high-Q optical or microwave cavity
\cite{Walls}. This opens up possibilities that go far beyond cooling
\cite{Hopkins03} and may not be within reach of other techniques for
manipulating a nano-resonator. In particular the technique presented
here would enable quantum state engineering of non-classical states of
motion \cite{Armour02,ion} along the lines proposed for the
trapped-ion system \cite{Poyatos96Law96}. Examples are the generation
and detection of a Fock state and squeezed states of motion.

\begin{figure}[!t]
\psfrag{0}{$ {\scriptstyle 10}^{\scriptscriptstyle -3}$}
\psfrag{8}{$ {\scriptstyle 10}^{\scriptscriptstyle -2}$}
\psfrag{2}{$ {\scriptstyle 10}^{\scriptscriptstyle -1}$}
\psfrag{3}{$\,\, {\scriptstyle 1}^{\scriptscriptstyle
    \vphantom 1}$}
\psfrag{4}{$ {\scriptstyle 10}^{\scriptscriptstyle
    \vphantom 1}$}
\psfrag{x}{$\ $}
\psfrag{y}{$\ $}
\psfrag{X}{$n_i$}
\psfrag{Y_}{$\!\!\! {}^{\textstyle \tilde n_f}$}
\psfrag{7}{$\!\!\!\!\! {\scriptstyle 10}^{\scriptscriptstyle -3}$}
\psfrag{5}{$\!\!\!\!\! {\scriptstyle 10}^{\scriptscriptstyle -2}$}
\psfrag{6}{$\!\!\!\!\! {\scriptstyle 10}^{\scriptscriptstyle -1}$}
\psfrag{g}{$\!\!\!\!\!\!\!\!\!\! {\scriptstyle  \left( \frac{\Gamma}{4
      \omega_0} \right) }^{\scriptscriptstyle 2}$}
\psfrag{G}{$\!\!\!\!\!\! {}^{{\scriptstyle
      \left( \frac{\Gamma}{4
      \omega_0} \right) }^{\scriptscriptstyle 2}}$}
\psfrag{a}{\scriptsize I}
\psfrag{b}{\scriptsize II}
\psfrag{c}{\scriptsize III}
\centerline{\includegraphics[width=8.5cm]{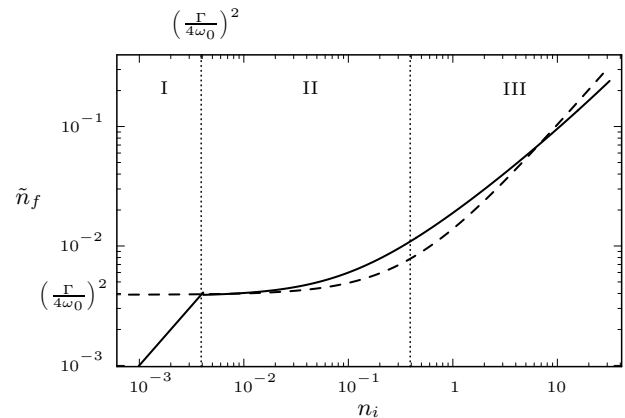}}
\caption{Optimal final number of phonons $\tilde n_f$ v.s.
initial number $n_i$ (solid line) for $\Gamma/\omega_0 = 1/4$,
$\eta^2 Q = 100$. The optimum results from searching for the laser
parameters ($\delta_{\mathrm{opt}}$, $\Omega_{\mathrm{opt}}$) that
minimize $n_f$. There are three regimes (dotted lines): I) for $n_i
\leq (\frac{\Gamma}{4 \omega_0})^2$ laser addressing of the structure
is not useful (i.e. $n_f \geq n_i$), II) for small $n_i$ ($n_i \ll
\eta^2 Q (\frac{\Gamma}{4 \omega_0})^2 $) the optimum is constant and
$\delta_{\mathrm{opt}} \approx \omega_0$, $\Omega_{\mathrm{opt}}^2 \ll
\omega_0^2$, III) for large $n_i$ ($n_i \gg \eta^2 Q (\frac{\Gamma}{4
\omega_0})^2 $) the optimum is proportional to $n_i$ with the ratio
$\tilde n_f/n_i$ bounded by $1/\eta^2 Q$ and $\delta_{\mathrm{opt}}
\sim \omega_0$, $\Omega_{\mathrm{opt}} \sim \omega_0$. The rough
approximation $n_i/\eta^2 Q + (\Gamma/4\omega_0)^2$ is
plotted for comparison (dashed line).}
\end{figure}

One could also envision extending this method to simultaneous cooling
of several modes. If successful, this could be used to generate
entangled states of mechanical motion. Our results could apply to
other systems with sharp vibrational resonances and to different types
of zero-dimensional emitters such as localized defects
\cite{Jelezko03}.

We thank A.~N.~Cleland for invaluable help and the KITP at the
University of California, Santa Barbara, for hospitality. We thank
R.~Blatt, J.~S.~Langer and Chiou-Fu Wang for helpful discussions. This
work was partially supported by a David \& Lucile Packard
Fellowship. Work at the University of Innsbruck is supported by the
Austrian Science Foundation, EU networks and the Institute for Quantum
Information.

\end{document}